\documentclass[preprint,showpacs,preprintnumbers,amsmath,amssymb]{revtex4}

\usepackage{graphicx,psfrag}% Include figure files
\usepackage{dcolumn}% Align table columns on decimal point
\usepackage{bm}% bold math
\begin{document}

\preprint{}

\title{Frequency Locking of an Optical Cavity using LQG Integral Control}% Force line breaks with \\

\author{S. Z. Sayed Hassen, M. Heurs, E. H. Huntington, I. R. Petersen} \affiliation{University of New South
  Wales at the Australian Defence Force Academy, School of Information Technology and Electrical Engineering,
  Canberra, ACT~2600, Australia} \email{s.sayedhassen@adfa.edu.au} \author{M. R. James}%
\affiliation{Department of Engineering, Australian National University, Canberra, ACT~2600, Australia}

\date{\today}% It is always \today, today,
% but any date may be explicitly specified


\begin{abstract}
  This paper considers the application of integral Linear Quadratic Gaussian (LQG) optimal control theory to a
  problem of cavity locking in quantum optics. The cavity locking problem involves controlling the error
  between the laser frequency and the resonant frequency of the cavity. A model for the cavity system, which
  comprises a piezo-electric actuator and an optical cavity is experimentally determined using a subspace
  identification method. An {LQG} controller which includes integral action is synthesized to stabilize the
  frequency of the cavity to the laser frequency and to reject low frequency noise. The controller is
  successfully implemented in the laboratory using a dSpace {DSP} board.
\end{abstract}

\maketitle

\section{Introduction}
\label{sec:intro}

Many future technologies will be based on quantum systems manipulated to achieve engineering outcomes \cite{MIL96,DM03}. Quantum feedback control forms one of the key design methodologies that will be required to achieve these quantum engineering objectives
 \cite{BEV83,MAB02,WM93a,HW95,AASDM02,RZK03,DJ99,HFCH04}. Examples of quantum systems in which quantum control may play a key role include the quantum error correction problem (see \cite{ADL02}) which is central to the development of a quantum computer and also important in the
problem of developing a repeater for quantum cryptography systems, spin control in coherent magnetometry (see
\cite{SGDM03}), control of an atom trapped in a cavity (see \cite{DJ99}), the control of a laser optical
quantum system (see \cite{YK03a}), control of atom lasers and Bose Einstein Condensates (see \cite{YJ08}), and
the feedback cooling of a nanomechanical resonator (see \cite{HJHS03}).

Attention is now turning to more general aspects of quantum control, particularly in the development of
systematic quantum control theories for quantum systems. For example in \cite{DJ99} and \cite{EB05} it was
shown that the linear quadratic Gaussian (LQG) optimal control approach to controller design can be extended
to linear quantum systems. Also, in \cite{JNP1}, it was shown that the $H^\infty$ optimal control approach to
controller design can be extended to linear quantum systems. These theoretical results indicate that
systematic optimal control methods of modern control theory have the potential of being applied to quantum
systems. Such modern control theory methods have the advantage that they are strongly model based and provide
systematic methods of designing multivariable control systems which can achieve excellent closed loop
performance and robustness. Experimental demonstrations of some of these theoretical results now appear
viable.  For example, Ref.~\cite{MAB08} presents the first experimental demonstration of the design and
implementation of a \lq\lq coherent controller\rq\rq \ from within this formalism.

One particular systematic approach to control is the LQG optimal control approach to design.  LQG optimal control
is based on a linear dynamical model of the plant being controlled which is subject to Gaussian white noise disturbances; e.g, see \cite{KS72}.   In LQG optimal control, a dynamic linear output feedback controller is sought to minimize a quadratic cost functional which encapsulates the performance requirements of the control system.
%Computationally tractable solutions to the LQG optimal control problem are available using
%standard software packages even in the case in which the model of the system being controlled is
%quite complex.
A feature of the LQG optimal control problem is that its solution involves the use of
a Kalman Filter which provides estimates of the internal system variables.  %LQG optimal control
%does not necessarily result in a controller which includes integral action. However,
Furthermore, in many applications integral action is required in order to overcome low frequency disturbances acting on the system being controlled. This issue is addressed here by using a version of LQG optimal control referred to as integral LQG control which forces the controller to include integral action; see \cite{G79}.

In this paper, we consider the application of systematic methods of LQG optimal control to the archetypal quantum optical problem of locking the resonant frequency of an optical cavity to that of a laser.  Homodyne detection of the reflected port of a Fabry-Perot cavity is used as the measurement signal for an integral LQG controller.  In our case, the linear dynamic model used is obtained using both physical considerations and experimentally measured frequency response data which is fitted to a linear dynamic model using subspace system identification methods; e.g., see \cite{MAL96}. The integral LQG controller design is discretized and implemented
on a dSpace digital signal processing (DSP) system in the laboratory and experimental results were obtained showing that the controller has been effective in locking the optical cavity to the laser frequency.  We also compare the step response obtained experimentally with the step response predicted using the identified model.

This paper is structured as follows:  In Section \ref{sec:problem} the quantum optical model of an empty cavity is formulated in a manner consistent with the LQG design methodology; Section \ref{sec:sysid-cavity} outlines the subspace system identification technique used to arrive at the linear dynamic model for the cavity system; Section \ref{sec:lqg} presents the LQG optimal controller design methodology as applied to the problem of locking the frequencies of a laser and of an empty cavity together; Section \ref{sec:expt} presents experimental results; and we conclude in Section \ref{sec:conclusion}.

\section{Cavity model}
\label{sec:problem}
\begin{figure}[h]
  \begin{center}
    \psfrag{bt}{$\tilde b$} \psfrag{mphi}{$-\phi$} \psfrag{bout}{$b_{out}$} \psfrag{bLout}{$b_{L,out}$}
    \psfrag{bL}{$b_L$} \psfrag{b1out}{$b_{1,out}$} \psfrag{b1}{$b_1$} \psfrag{bnt}{$b$}
    \psfrag{y}{\hspace{10pt}$y$} \psfrag{u}{\hspace{5pt}$u$}
    \includegraphics[width=8.0cm]{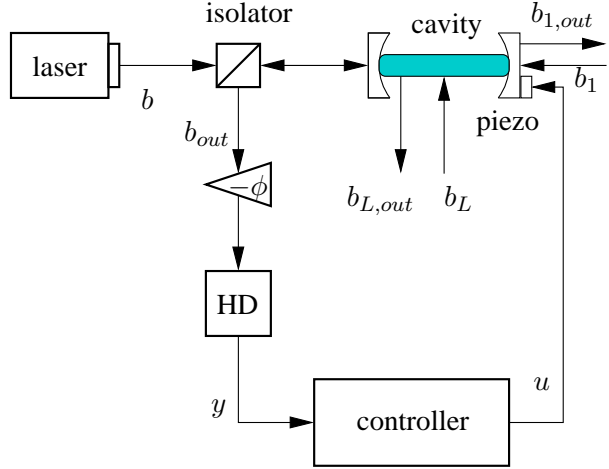}
    \caption{Cavity locking feedback control loop.}
    \label{fig:hjp-laser1-fig1}
  \end{center}
\end{figure}

A schematic of the frequency stabilization system is depicted in Fig.~\ref{fig:hjp-laser1-fig1}.

The cavity can be described in the Heisenberg picture by the following quantum stochastic differential
equations; e.g., see  \cite{BR04} and Section 9.2.4 of \cite{GZ00}:
\begin{eqnarray}
  \dot a &=& -(  \frac{\kappa}{2} + i \Delta)a - \sqrt{\kappa}_0 (\beta
  + b_0) \nonumber \\
  && \quad - \sqrt{\kappa}_1 b_1 - \sqrt{\kappa}_L b_L ;
  \nonumber \\
  b_{out} &=& \sqrt{\kappa}_0 a + \beta+ b_0.
  \label{cavity-1}
\end{eqnarray}
Here, the annihilation operator for the cavity mode is denoted by $a$
and the annihilation operator for the coherent input mode is denoted by
$b=\beta+b_0$, both defined in an appropriate rotating  
reference frame. Here $b_0$ is quantum noise.  We have written $ \kappa = \kappa_0 + \kappa_1 + \kappa_L , $ where $\kappa_0$, $\kappa_1$
and $\kappa_L$ quantify the strength of the couplings of the respective optical fields to the cavity,
including the losses.  The input to the cavity is taken to be a coherent state with amplitude $\beta$ and which
is assumed to be real without loss of generality. $\Delta$ denotes the frequency detuning between the laser
frequency and the resonant frequency of the cavity. The objective of the frequency stabilization scheme is to
maintain $\Delta=0$. The detuning is given by \cite{siegman}
\begin{equation}
  \Delta = \omega_c - \omega_L = q \frac{2\pi c}{nL} - \omega_L,
\end{equation}
where $\omega_c$ is the resonant frequency of the cavity, $\omega_L$ is the laser frequency, $nL$ is the
optical path length of the cavity, $c$ is the speed of light in a vacuum and $q$ is a large integer indicating
that the $q^{\mathrm{th}}$ longitudinal cavity mode is being excited.

The cavity locking problem is formally a nonlinear control problem
since the equations governing the cavity
dynamics in (\ref{cavity-1}) contain the nonlinear product term $\Delta\, a$. In order to apply linear
optimal control techniques, we linearize these equations about the zero-detuning point.  Let $\alpha$ denote
the steady state average of $a$ when $\Delta=0$ such that $ a = \alpha + \tilde a$.  The perturbation operator
$\tilde a$ satisfies the linear quantum stochastic differential equation (neglecting higher order terms)
\begin{equation}
  \dot{\tilde{a}} = -\frac{\kappa}{2} \tilde a -i \Delta \alpha - \sqrt{\kappa}_0 b_0 -
  \sqrt{\kappa}_1 b_1 - \sqrt{\kappa}_L b_L. \label{tilde-a-2}
\end{equation}
The perturbed output field operator $\tilde b_{out}$ is given by
\begin{equation}
  \tilde b_{out}= \sqrt{\kappa}_0 \tilde a + b_0 \label{tilde-b-out}
\end{equation}
which implies $b_{out}= \sqrt{\kappa}_0 \alpha +\beta + \tilde b_{out}$.

We model the measurement of the $X_\phi$ quadrature of $\tilde b_{out}$ with homodyne detection by changing
the coupling operator for the laser mode to $\sqrt{\kappa}_0 e^{-i \phi} a$, and measuring the real quadrature
of the resulting field. The measurement signal is then given by
\begin{eqnarray}
  \tilde y &=&  \tilde b_{out} + \tilde b_{out}^\dagger
  % \nonumber \\
  = \sqrt{\kappa}_0( e^{-i \phi} \tilde a + e^{i \phi} \tilde a^\dagger ) + q_0 \label{y-eqn}
\end{eqnarray}
where $ q_0 = b_0 + b_0^\dagger $ is the intensity noise of the input coherent state.

As we shall see in Section \ref{sec:lqg}, the LQG controller design process starts from a state-space model of
the plant, actuator and measurement, traditionally expressed in the form:

\begin{eqnarray}
  \dot{x} &=& Ax + Bu+D_1w; \nonumber \\
  y &=& Cx+D_2 w \label{eqn:cavity-sys}
\end{eqnarray}
where $x$ represents the vector of system variables (we shall not use the control engineering term ``states'' to
avoid confusion with quantum mechanical states), $u$ is the input to
the system, $y$ is the measured output, and $w$ is a Gaussian white
noise disturbance acting on the system. Also, $A$ represents the
system matrix, $B$ is the input matrix, $C$ is the output matrix, and
$D_1$ and $D_2$ are the system noise matrices.

The cavity dynamics and homodyne measurement are expressed in state-space form in terms of the quadratures of
the operators $\tilde{a}, b_0, b_1, b_L$ as
\begin{eqnarray}
  \left[  \begin{array}{c} \dot{\tilde{q}} \\  \dot{\tilde{p}} \end{array} \right] &=& \left[
    \begin{array}{cc} -\frac{\kappa}{2} & 0 \\ 0 & -\frac{\kappa}{2} \end{array} \right] \left[
    \begin{array}{c} \tilde{q} \\  \tilde{p} \end{array} \right] + \left[   \begin{array}{c} 0 \\
      2\alpha \end{array} \right] \Delta \nonumber \\ && - \sqrt{\kappa}_0 \left[   \begin{array}{cc}
      \cos \phi & -\sin \phi \\ \sin \phi & \cos \phi \end{array} \right]
  \left[ \begin{array}{c} q_0 \\  p_0 \end{array} \right]
  \nonumber \\ && - \sqrt{\kappa}_1 \left[   \begin{array}{cc} 1 & 0 \\ 0 & 1 \end{array} \right]
  \left[ \begin{array}{c} q_1 \\  p_1 \end{array} \right] \nonumber \\ && - \sqrt{\kappa}_L \left[
    \begin{array}{cc} 1 & 0 \\ 0 & 1 \end{array} \right] \left[ \begin{array}{c} q_L \\  p_L
    \end{array} \right];
  \nonumber \\
  y &=& k_2 \sqrt{\kappa}_0 \left[   \begin{array}{cc} \cos \phi & \sin \phi \end{array} \right]
  \left[  \begin{array}{c} \tilde{q} \\  \tilde{p} \end{array} \right] \nonumber \\ && + \; k_2\left[
    \begin{array}{cc} 1  &  0 \end{array} \right] \left[  \begin{array}{c} q_0  \\  p_0 \end{array}
  \right] + w_2 \nonumber \\
&=& z+  \; k_2\left[
    \begin{array}{cc} 1  &  0 \end{array} \right] \left[  \begin{array}{c} q_0  \\  p_0 \end{array}
  \right] + w_2
\label{cavity-quad-linear} 
\end{eqnarray}
with noise quadratures $ q_j = b_j + b_j^\dagger, \ \ p_j = -i(q_j - p_j^\dagger),$ for $j=0,1,L$ (all
standard Gaussian white noises).  Here, $y$ is the homodyne detector output in which we have included an
electronic noise term $w_2$. Also, $k_2$ represents the transimpedance gain of the homodyne detector,
including the photodetector quantum efficiency.

\section{Subspace system identification}
\label{sec:sysid-cavity}

The state-space model of the cavity given in (\ref{cavity-quad-linear}) is incomplete as it does not include
an explicit model, including the actuation mechanism, for the dynamics of the detuning, $\Delta$.  The
dynamics of the detuning and actuation mechanism are sufficiently complex that direct measurement is a more
experimentally reasonable approach than {\it a priori} modeling of the system.  Following these measurements,
an approach called subspace system identification is used to obtain the complete state-space model.

It is at this point that the controller design process diverges from traditional, pre-1960s control
techniques.  Specifically, in the traditional approach, a controller is designed using root-locus or frequency
response methods, based on measurements of the plant transfer function \cite{Dorf}.  In our modern control
approach, the subspace identification method determines a state-space model from the input-output frequency
response data and generates the system matrices $A$, $B$ and $C$ in (\ref{eqn:cavity-sys}); see~\cite{OM96}.  The
system matrices are used in the LQG design process as we shall see in Section \ref{sec:lqg}.

% Physically, the dynamics of the detuning are governed by the dynamics of the laser frequency and by the
% dynamics of the cavity's optical path length.  For convenience we treat the dynamics of the optical path
% length as arising solely from the dynamics of the physical path length, which comprises mechanical
% disturbances as well as the actuation mechanism, achieved through a piezo-electric transducer (PZT).  For the
% purposes of describing the experiment conducted here, we assume no actuation on the laser frequency.

\begin{figure}[htb]
  \begin{center}
    \includegraphics[width=8.7cm]{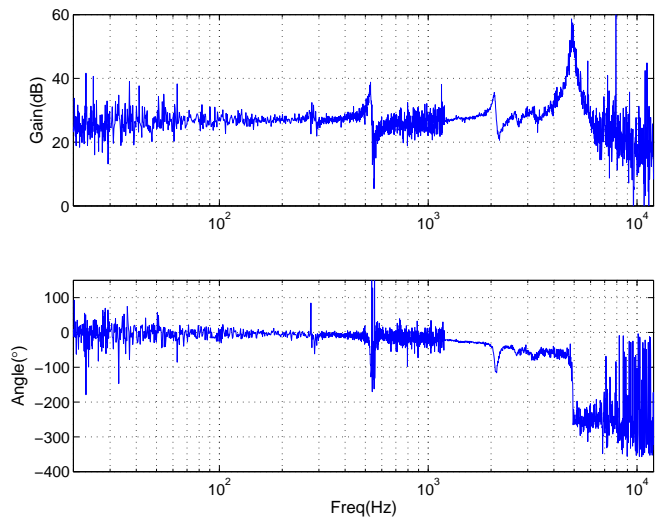}
  \end{center}
  \caption{Measured plant frequency response.}
  \label{fig:plant_sysid}
\end{figure}

The transfer function of the cavity and measurement system (or alternatively the ``plant'') is identified under
closed-loop conditions.  We use an analog proportional-integral (PI) controller to stabilize the system for
the duration of the measurement. The frequency response data thus obtained for the plant is plotted in
Fig.~\ref{fig:plant_sysid}.  From the data, at least three resonances can be clearly identified, occurring at
frequencies of about 520, 2100 and 5000 Hz respectively.

% Plot generated by system_id.m

An $8^\mathrm{th}$-order anti-aliasing filter with a corner frequency of 2.5 kHz (chosen because it is far
greater than the unity-gain bandwidth of the controller and far less than the 50 kHz sampling frequency) was
placed immediately prior to the digital LQG controller.  This is treated mathematically as augmenting the
plant such that the augmented plant has a frequency response that is the product of the plant data gathered
previously and the anti-aliasing filter which is identified separately.

\begin{figure}[htb]
  \begin{center}
    \includegraphics[width=8.7cm]{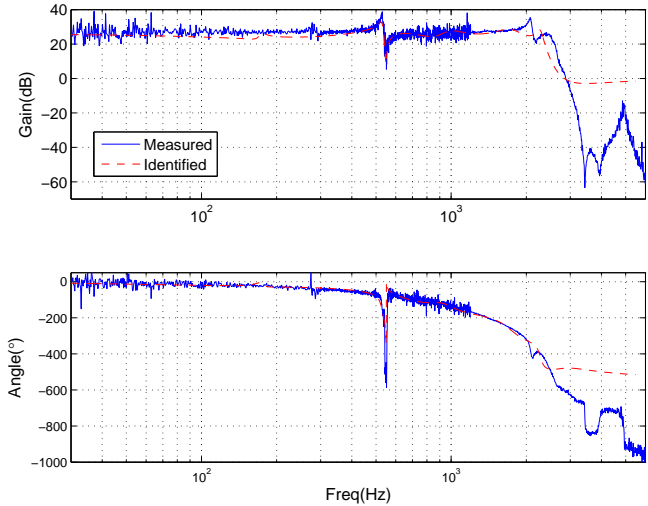}
  \end{center}
  \caption{Measured frequency response data for the augmented plant and frequency response for the identified
    system model.}
  \label{fig:hjp-laser1-sysid}
\end{figure}

The frequency response data obtained is then fitted to a $13^\mathrm{th}$ order model using a
subspace identification method; see \cite{MAL96}. The algorithm accommodates arbitrary frequency spacing
and is known to provide good results with flexible structures.  This makes it suitable for our
application which includes a piezo-electric actuator coupled to the cavity
mirror. Fig.~\ref{fig:hjp-laser1-sysid} compares the gain (in dB) and the phase (in $\deg$) of the
measured frequency data for the augmented plant with that of its identified system model.

% Plot generated by running system_id.m

\section{Linear Quadratic Gaussian Optimal Control}
\label{sec:lqg}
The LQG optimal control approach to controller design begins with a
linear state-space model of the form (\ref{eqn:cavity-sys}). Note that this model involves the use of Gaussian white noise
disturbances although a more rigorous formulation of the LQG optimal
control problem involves the use of a Wiener process to  describe the
noise rather than the white noise model (\ref{eqn:cavity-sys}); e.g.,
see \cite{AST70}. However, for
the purposes of this paper, a model of the form (\ref{eqn:cavity-sys})
is most convenient. In the model (\ref{eqn:cavity-sys}), the term $D_1
w$ corresponds to the process noise and the term $D_2 w$ corresponds
to the measurement noise. The LQG optimal control problem involves
constructing a dynamic measurement feedback controller to minimize a
quadratic cost functional of the form
\begin{equation}
  \label{eq:lqr}
 \mathcal{J} =   \lim_{T \rightarrow \infty}\mathbf{E} \left[
    \frac{1}{T} \int_0^T [ x^T Q x + u^T R u ] dt
  \right]
\end{equation}
where $Q \geq 0$ and $ R > 0$ are  symmetric weighting matrices. The
term $x^T Q x$ in the cost functional (\ref{eq:lqr}) corresponds to a
requirement to minimize the system variables of interest and the term $u^T R u$
corresponds to a requirement to minimize the size of the  control
inputs. The matrices $Q$ and $R$ are chosen so that the cost
functional reflects the desired performance objectives of the control
system. The great advantage of the LQG optimal control approach to
controller design is that it provides a tractable systematic way to construct
output feedback controllers (even in the case of multi-input
multi-output control systems). Also,  numerical solutions
exist in terms of algebraic Riccati equations which can be solved
using standard software packages such as Matlab; e.g., see
\cite{AM90,KS72}. A feature of the solution to the LQG optimal control
problem is that it involves a Kalman filter which provides an optimal
estimate $\hat x$ of the vector of system variables $x$ based on the
measured output $y$. This is combined with a ``state-feedback''
optimal control law which is obtained by minimizing the cost
functional (\ref{eq:lqr}) as if the vector of system variables $x$ was
available to the controller.

Note that the LQG controller design methodology cannot
directly handle some important engineering issues in control system design such as robustness margins and
controller bandwidth. These issues can however be taken into account in the controller design process by adding
extra noise terms to the plant model (over and above the noise that is present in the physical system) and by
suitably choosing the quadratic cost functional (\ref{eq:lqr}).

\subsection{LQG Performance Criterion and Integral Action}
\label{sec:lqg-optimal}

\begin{figure}[htb]
  \begin{center}
    \psfrag{D}{\hspace{-5pt}$\Delta$} \psfrag{u}{$u$} \psfrag{y}{$y$} \psfrag{w2}{$w_2$}
    \psfrag{w1}{$w_1$}%\psfrag{Quantum Noises}{\hspace{-5pt} Quantum Noises}
    \includegraphics[width=8.7cm]{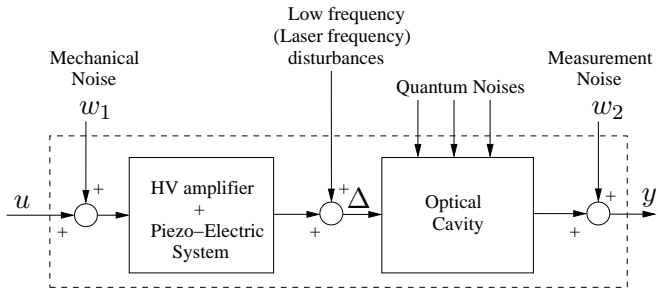}
    \caption{Block diagram of the plant.}
    \label{fig:sys}
  \end{center}
\end{figure}

The dynamics of the cavity and measurement system can be subdivided as shown in Fig.~\ref{fig:sys} and comprises an electro-mechanical subsystem and an electro-optical subsystem (the optical cavity and homodyne detector).

The control objective is to minimize the cavity detuning $\Delta$, which is not available for
measurement. Instead, the measurement signal $y$ is the output of the homodyne detector and to
include $\Delta$ in the performance criterion, we need to relate
$\Delta$ to $z$, the mean value of $y$. It can be seen
from (\ref{cavity-quad-linear}) that the transfer function of the optical cavity from $\Delta$ to
$z$ is a first-order low-pass filter with a corner frequency of $\kappa/2$.  Physically, this arises
from the well-known (see for example~\cite{siegman,Cheng} and the references therein) transfer
function of the optical cavity from $\Delta$ to a phase shift, which is then measured by the
homodyne detector.  In the experimental system described herein $\kappa/2\approx 10^5$~Hz, which is
well beyond the frequency range of interest for the integral {LQG} controller. Hence we can consider
$z$ to be proportional to $\Delta$ under these conditions, and therefore minimizing variations in
$\Delta$ can be regarded as being equivalent to minimizing variations in $z$.

The LQG performance criterion to be used for our problem is chosen to reflect the desired control system
performance. That is, (i) to keep the cavity detuning $\Delta$ small (ideally zero), and (ii) to
limit the control energy. However, these requirements are not sufficient to generate a suitable
controller as the system is subject to a large initial DC offset and slowly varying disturbances.
Our application requires the elimination of such effects. This
can be achieved by using integral action and is the reason for our use of the integral LQG
controller design method.

We include integral action by adding an additional term in the cost function which involves the
integral of the quantity $z$. Moreover, we include the ``integral state'' as another  variable of
the system.  The new variable $\int z\,dt$ is also fed to the Kalman filter, which when combined with
an optimal state-feedback control law leads to an integral {LQG}
optimal controller. This controller
will then meet the desired performance requirements as described above; e.g., see~\cite{G79}.

\begin{figure}[htb]
  \begin{center}
    \psfrag{int}{$\;\int$} \psfrag{w1}{\hspace{-10pt}$w_1$} \psfrag{w2}{\hspace{4pt}$w_2$} \psfrag{w3}{$w_3$}
    \psfrag{K}{\hspace{-11pt}Controller} \psfrag{u}{$u$}
    \psfrag{y}{$z$} \psfrag{z1}{$y_1$} \psfrag{z2}{$y_2$} 
    \psfrag{PLANT}{\hspace{10pt}Plant}
    \includegraphics[width=7cm]{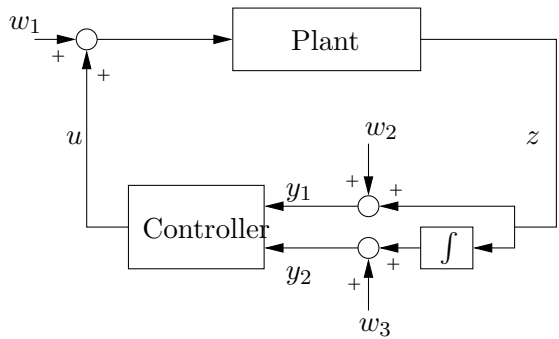}
  \end{center}
  \caption{ Integral {LQG} controller design configuration}
  \label{fig:sysloop}
\end{figure}

Fig.~\ref{fig:sysloop} shows the integral {LQG}  controller design configuration.

The overall system can be described in state-space form as follows:
\begin{eqnarray}
  \dot{\tilde{x}} &=& \tilde{A} \tilde{x} + \tilde{B} w_1 + \tilde{B} u; \nonumber \\
  \tilde{y} &=& \tilde{C} \tilde{x} + \left[ \begin{array}{c} w_2 \\ w_3 \end{array} \right];
  \label{sysid}
\end{eqnarray}
where $$ \tilde{x} = \left[ \begin{array}{c} x \\ \int z\,d\tau \end{array} \right] \;\;\mathrm{and}\;\;
\tilde{y} = \left[ \begin{array}{c} y_1 \\ y_2
  \end{array} \right]. $$
Here the
matrices $\tilde{A}, \tilde{B}, \tilde{C}$ are constructed from the matrices $A, B, C$ as follows:
$$
\tilde{A} = \left[ \begin{array}{cc} A & 0 \\C & 0 \end{array} \right], \quad \tilde{B} = \left[
  \begin{array}{c} B \\ 0 \end{array} \right],\; \mathrm{and} \;\; \tilde{C} = \left[ \begin{array}{cc} C &
    0\\0 & I\end{array} \right].
$$

Section \ref{sec:sysid-cavity} outlines the technique used to determine $A, B, C$.

In equation~(\ref{sysid}), the quantity $w_1$ represents mechanical noise entering the system which
is assumed to be Gaussian white noise with variance $\epsilon_1^2$. The quantity $w_2$ represents
the sensor noise present in the system output $y$, which is assumed to be Gaussian white noise with
variance $\epsilon_2^2$. The quantity $w_3$ is included to represent
the sensor noise added to  the quantity
 $\int z\,dt$. This is assumed to be Gaussian white noise with variance $\epsilon_3^2$ and is
included to fit into the standard framework for the {LQG} controller design. The parameters
$\epsilon_1, \epsilon_2$ and $\epsilon_3$ are treated as design parameters in the {LQG} controller
design in Sec.~\ref{sec:lqg-values}.

The integral {LQG} performance criterion can be written as:
\begin{equation}
  \mathcal{J} =   \lim_{T \rightarrow \infty}\mathbf{E} \left[
    \frac{1}{T} \int_0^T [ x^T Q x + L(z)^T \bar{Q} L(z) + u^T R u ] dt
  \right] \label{LQG-cost}
\end{equation}
where
$$ L(z) = \int\nolimits_0^t z(\tau)\, d\tau. $$
We choose the matrices $Q, R$ and $\bar{Q}$ such that
$$  x^T Q x = \vert z\vert^2,  \ \ u^T R u = r \vert u \vert^2 ,\,
\mathrm{and}\;\; \bar{Q} = \bar{q}, $$ where $r > 0$ and $\bar{q} > 0$ are also treated as design parameters.

The first term of the integrand in (\ref{LQG-cost}) ensures that the controlled variable $z$ goes to
zero, while the second term forces the integral of the controlled variable to go to zero. Also, the
third term serves to limit the control input magnitude. The expectation in (\ref{LQG-cost}) is with
respect to the classical Gaussian noise processes described previously, and the assumed Gaussian
initial conditions. Given our system as described by (\ref{sysid}), the optimal {LQG} controller is
given by (e.g., see \cite{KS72})
\begin{eqnarray}
  u &=& - r^{-1} \tilde{B}^T X \hat{\tilde{x}} = F\hat{\tilde{x}}, \label{LQG-u}
\end{eqnarray}
where $X$ is the solution of the following matrix Riccati equation
\begin{eqnarray}
  0 &=& X\tilde{A} + \tilde{A}^TX + \tilde{Q}-r^{-1}X \tilde{B}^T\tilde{B}X, \label{LQG-X}
\end{eqnarray}
and
\begin{eqnarray*}
  \tilde{Q} = \tilde{C}^T \left[ \begin{array}{cc} 1 & 0 \\ 0 & \bar{q} \end{array} \right]
  \tilde{C}.
\end{eqnarray*}
Here $\hat{\tilde{x}}$ is an optimal estimate of the vector of plant
variables $\tilde x$ obtained via a steady state Kalman filter which
can be described by the state equations
\begin{equation}
 \dot{\hat{\tilde{x}}} = \tilde{A} \hat{\tilde{x}} + \tilde{B}u + K
  [\tilde{y}-\tilde{C}\hat{\tilde{x}}].
  \label{LQG-filter}
\end{equation}
For the case of uncorrelated process and measurement noises, the steady state Kalman filter is obtained by
choosing the gain matrix $K$ in (\ref{LQG-filter}) as
\begin{eqnarray}
  K &=& P \tilde{C}^T V_2^{-1}, \label{LQG-Y}
\end{eqnarray}
where $P$ is the solution of the matrix Riccati equation
\begin{eqnarray}
  0 &=& \tilde{A}P + P\tilde{A}^{T} + V_1 - P\tilde{C}^TV_2^{-1}\tilde{C}P.
\end{eqnarray}
Here
\begin{eqnarray*}
  V_1 = \epsilon_1^2 \tilde{B} \tilde{B}^T = \mathbf{E}[w_1 w_1^T] \quad \mathrm{and} \quad V_2 =
  \left[
    \begin{array}{cc} \epsilon_2^2 & 0 \\ 0 & \epsilon_3^2
    \end{array} \right]
\end{eqnarray*}
define the covariance of the process and measurement noises respectively.

\subsection{Design Parameters}
\label{sec:lqg-values}

In designing the LQG controller, the parameters $\epsilon_1^2$ (the mechanical noise
variance), $\epsilon_2^2$ (the sensor noise variance of $y$),
$\epsilon_3^2$ (the variance of the sensor noise added to  
 $\int z$), $r$ (the control energy weighting in the LQG cost function) and $\bar{q}$
(the integral output weighting in the LQG cost function) were used as design parameters and were adjusted for good controller
performance. This includes a requirement that the control system have suitable gain and phase margins and a reasonable controller bandwidth. The specific values used for the design are shown in Table~\ref{tab:designvalues}:
\begin{table}[htbp]
  \begin{center}
    \begin{tabular}{||c|l||}
      \hline Design parameter & Value \\ \hline \hline
      $\epsilon_1$ & $5 \times 10^{\scriptscriptstyle -2}$ \\
      $\epsilon_2$ & $500$ \\
      $\epsilon_3$ & $3 \times 10^{\scriptscriptstyle -4}$ \\
      $r$ & $1 \times 10^{\scriptscriptstyle 3}$ \\
      $\bar{q}$ & $1 \times 10^{\scriptscriptstyle 6}$ \\ \hline
    \end{tabular}
  \end{center}
  \caption{Design Parameter Values}
  \label{tab:designvalues}
\end{table}

\begin{figure}[htb]
  \begin{center}
    \includegraphics[width=8.7cm]{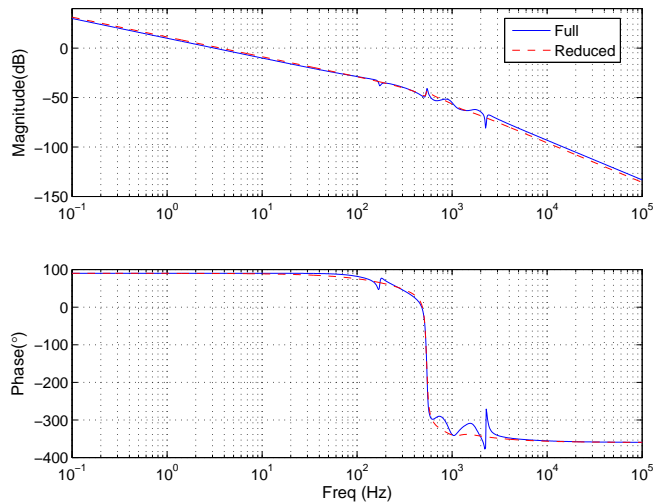}
  \end{center}
  \caption{Full and reduced-order continuous LQG controller Bode plots.}
  \label{fig:cont_fr}
\end{figure}

These parameter values led to a $15^\mathrm{th}$ order {LQG} controller which is reduced to a $6^\mathrm{th}$
order controller using a frequency-weighted balanced controller reduction approach. We reduce the order of the
controller to decrease the computational burden (and hence time-delay) of the controller when it is discretized
and implemented on a digital computer. The new lower order model for
the controller is determined via a certain controller reduction
technique which minimizes the weighted frequency response error
between the original controller transfer function and the reduced
controller transfer function; see \cite{ZDG96}. This is illustrated 
 in Fig.~\ref{fig:cont_fr} which shows Bode plots of the  full-order
controller and the reduced order controller. Formally, the
reduced order controller is constructed so that the quantity
\begin{equation}
  \left \| \frac{P(s)C(s)}{(I+P(s)C(s))} (C(s) - C_r(s)) \right \|_\infty \label{eqn:cont_red_infnorm}
\end{equation}
is minimized.  Here, $P(s) =  \tilde C(sI-\tilde
A)^{-1}\tilde B$, is the
plant transfer function matrix and $C(s) = F(sI-\tilde A + K \tilde C
- \tilde B F)^{-1}K$ is the original full controller transfer function
matrix. Also, $C_r(s)$ is the reduced dimension controller transfer
function matrix. The notation $\|M(s)\|_\infty$ refers to the $H^\infty$
norm of a transfer function matrix $M(s)$ which is defined to be the
maximum  of $\sigma_{max}[M(i\omega)]$ over all frequencies $\omega
\geq 0$. Here
$\sigma_{max}[M(i\omega)]$ refers to the maximum singular value of the
matrix $M(i\omega)$.

Note that this approach to controller reduction  does not guarantee
the stability of the closed loop system with the  reduced
controller. We check for closed loop stability
stability separately after the controller reduction process.  

% Figure generated by running laser_lqgdesign.m

\begin{figure}[htb]
  \begin{center}
    \includegraphics[width=8.7cm]{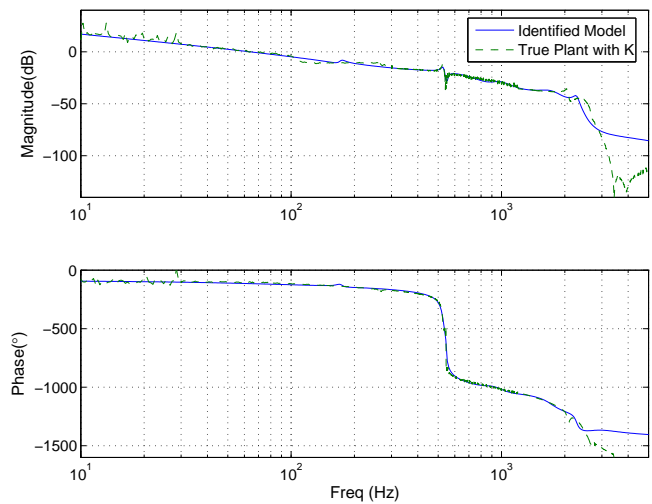}
  \end{center}
  \caption{Bode plot of loop gain $L$. The controller provides a gain margin of 16 dB and $63^{\circ}$ phase margin.}
  \label{fig:Loop_Gain}
\end{figure}

The reduced controller is then discretized at a sampling rate of 50 kHz and the corresponding Bode plot of the
discrete-time loop gain is shown in Fig. \ref{fig:Loop_Gain}. This
discretized controller provides good gain and phase
margins of 16.2 dB and $63^{\circ}$ respectively.
% Figure generated by running laser_lqgdesign.m

\section{Experimental Results}
\label{sec:expt}
\begin{figure}[htb]
  \begin{center}
    \includegraphics[width=8.7cm]{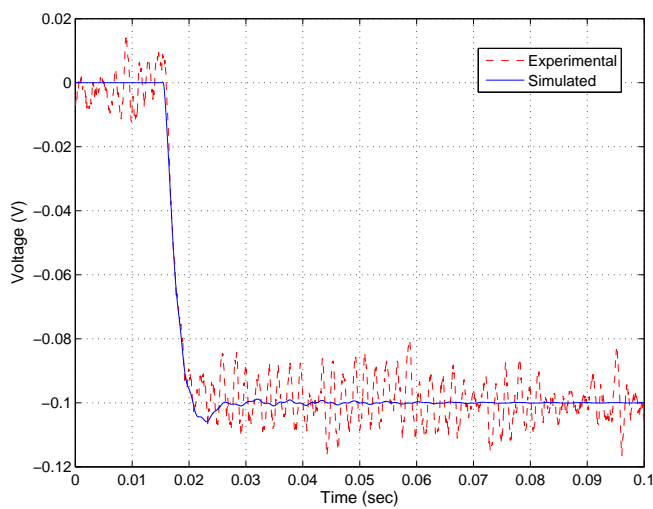}
  \end{center}
  \caption{Step Response of the closed-loop system to an input of 0.1 V.}
  \label{fig:stepresp}
\end{figure}

\begin{figure}[htb]
  \begin{center}
    \psfrag{r}{$r$} \psfrag{e}{$e$} \psfrag{y}{$y$} \psfrag{z2}{$z_2$} \psfrag{PLANT}{\hspace{5pt}PLANT}
    \psfrag{Controller}{\hspace{5pt}Controller}
    \includegraphics[width=5cm]{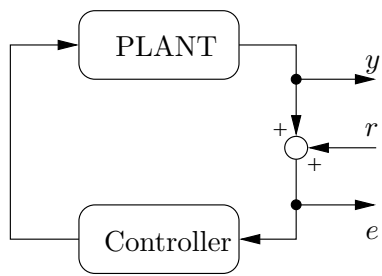}
  \end{center}
  \caption{Setup used to measure the closed-loop step response}
  \label{fig:e_to_r}
\end{figure}

The discrete controller is implemented on a {dSpace} {DS1103} Power PC
DSP Board. This board is fully
programmable from a Simulink block diagram and possesses 16-bit
resolution. The controller successfully stabilizes the
frequency in the optical cavity, locking its resonance to that of the laser frequency, $f_0$; see
\cite{SHJP1a}. This can be seen from the experimentally measured step
response shown in  Fig.~\ref{fig:stepresp} This step response was
measured by applying a step disturbance of magnitude 0.1~V  to the
closed-loop system as shown in Figure \ref{fig:e_to_r}. Here $r$ is the step input
signal and $y$ is the resulting step response signal which was measured.  

% Figure generated by running compare_step.m

\section{Conclusion and Future Work}
\label{sec:conclusion}

In this paper, we have shown that a systematic modern control technique such as {LQG} integral control can be applied to a problem in experimental quantum optics which has previously been addressed using traditional approaches to controller design. From frequency response data gathered, we have successfully modeled the optical cavity system, and used an extended version of the {LQG} cost functional to formulate the specific requirements of the control problem. A controller was obtained and implemented which locks the resonant frequency of the cavity to that of the laser frequency.

To improve on the current system, one might consider using additional actuators such as a phase modulator situated within the cavity or an additional piezo actuator to control the driving laser. Additional sensors which could be considered include using a beam splitter and another homodyne detector to measure the other optical quadrature and an accelerometer or a capacitive sensor to provide additional measurements of the mechanical subsystem. In addition, it may be useful to control the effects of air turbulence within the cavity and an additional interferometric sensor could be included to measure the optical path length adjacent to the
cavity. Such a measurement would be correlated to the air turbulence effects within the cavity. All of these additional actuators and sensors could be expected to improve the control system performance provided they were appropriately exploited using a systematic multivariable control system design methodology such as LQG control.

One important advantage of the {LQG} technique is that it can be extended in a straightforward way to multivariable control systems with multiple sensors and actuators. Moreover, the subspace approach to identification used to determine the plant is particularly suited to multivariable systems. It is our intention to further the current work by controlling the laser pump power using a similar scheme as the one used in this paper. This work is expected to pave the way for extremely stable lasers with fluctuations approaching the quantum noise limit and which could be potentially used in a wide range of applications in high precision metrology, see \cite{Huntington07}.

%\bibliography{pra}% Produces the bibliography via BibTeX.
%\bibliography{/home/irp/Bibliog/irpnew.bib}
%\bibliography{/home/sayed/Documents/PhD_Work/Bibliography/irpnew.bib}
%\bibliography{irpnew}

\end{document}